\RequirePackage{fix-cm}
\documentclass[smallextended]{svjour3}       
\smartqed  
\usepackage{graphicx}
\usepackage{bm}
\usepackage{hyperref}
\usepackage{natbib}
\usepackage{braket}
\usepackage{float}
\begin{document}

\title{A Modified $8f$ Geometry With Reduced Optical Aberrations For Improved Time Domain Terahertz Spectroscopy}


\titlerunning{A Modified $8f$ Geometry With Reduced Optical Aberrations For TDTS}        

\author{N. J. Laurita \and Bing Cheng \and R. Barkhouser \and V. A. Neumann \and N. P. Armitage}


\institute{N. J. Laurita \at 
				The Institute for Quantum Matter, Department of Physics and Astronomy, The Johns Hopkins University, Baltimore, MD 21218, USA \\
              \email{Laurita.Nicholas@Gmail.com}           
          		 \and
           Bing Cheng \at
               The Institute for Quantum Matter, Department of Physics and Astronomy, The Johns Hopkins University, Baltimore, MD 21218, USA
               \and
           R. Barkhouser \at
               Instrument Development Group, Department of Physics and Astronomy, The Johns Hopkins University, Baltimore, MD 21218, USA
               \and
               V. A. Neumann \at
               The Institute for Quantum Matter, Department of Physics and Astronomy, The Johns Hopkins University, Baltimore, MD 21218, USA \at Faculty of Science \& Technology, University of Twente, 7500 AE Enschede, The Netherlands
               \and
               N. P. Armitage \at The Institute for Quantum Matter, Department of Physics and Astronomy, The Johns Hopkins University, Baltimore, MD 21218, USA 
}

\date{Received: date / Accepted: date}

\maketitle

\begin{abstract}
\noindent We present a modified $8f$ geometry for time domain terahertz (THz) spectroscopy (TDTS) experiments.  We show, through simulations and data, that a simple rearranging of the off-axis parabolic mirrors, which are typically used to focus and direct THz radiation in TDTS experiments, results in a nearly 40\% reduction in the THz focal spot diameter.  This effect stems from significant reduction of the principle optical aberrations which are \textit{enhanced} in the conventional $8f$ geometry but partially \textit{compensated} in the modified $8f$ experimental setup.  We compare data from our home-built TDTS spectrometer in the modified $8f$ geometry to that of previous iterations that were designed in the conventional $8f$ geometry to demonstrate the effect.

\end{abstract}

\section{Introduction}
\label{intro}

Recent advances in terahertz (THz) generation and detection have resulted in a plethora of techniques which probe picosecond time scales with exceptionally high precision. Among these techniques, time domain terahertz spectroscopy (TDTS) \cite{Nuss1998, Tonouchi2007} has evolved into a ubiquitous tool within both the commercial and academic research landscape.  Industrially, TDTS has been utilized in the fields of pharmaceuticals \cite{Taday351}, medical imaging \cite{Hu:95, Johnson2001, Mittleman:97, Wang2003, Chan2007}, protein detection \cite{Nagel:02}, and biohazard applications \cite{Choi337, LeahyHoppa2007227}. TDTS has also made significant contributions to active fields of basic research such as materials characterization \cite{Kaindl2003, Heyman1998}, superconductivity \cite{Corson1999, Bilbro2011}, quantum magnetism \cite{Morris2014, Laurita2015, Pan2015}, and topological insulators \cite{Aguilar2012, Wu2013, Hancock2011}.  However, despite significant progress in the development of THz spectroscopy, limiting factors remain.  Perhaps the largest constraint is the technique's requirement for exceptionally large sample sizes resulting from large THz focal spots at the sample position.  While large focal spots are partially caused by the difficulty of tightly focusing long wavelength THz radiation, 1 THz $=$ 0.3 mm, the observed focal spot sizes of up to 10 mm in conventional systems are far larger than the wavelength would suggest. Thus, a need exists to better understand THz pulse propagation through such TDTS spectrometers and improve their imaging capabilities.

The advent of THz cameras have greatly aided in the efforts to optimally fine tune experimental set-ups to obtain diffraction limited THz pulses.  Although camera sensitivity limits their use to only high intensity THz experiments, the profile of THz pulses have been three-dimensionally real space imaged from a variety of such high intensity sources such as LiNbO$_3$ \cite{Hirori2011}, two-color air plasmas \cite{Klarskov2013}, and organic nonlinear crystals \cite{Shalaby2014}.  In these cases, the beam profile is able to be monitored while tuning some experimental parameter to optimize system performance.  However, obtaining optimal performance for spectrometers based on the photoconductive antenna method of generating and detecting THz radiation, as is the case in this work, is significantly more difficult as the generated THz electric field lies below the detection threshold of such cameras.  In principle, one can raster scan the detector to image THz pulses, which has been done with success \cite{Reiten2003}, but doing so is time consuming and therefore not a practical method of fine tuning spectrometers.  Instead, new methods for improving the performance of low intensity spectrometers need to be developed in order to reduce the constraint imposed by sample size on the technique.

Recently, we demonstrated that smaller THz focal spots can be achieved via a nested coaxial parabolic reflector placed at the sample position \cite{Neumann2015}.  In the present work, we show that reduced focal spot sizes can also be achieved via a particular arrangement of the off axis parabolic mirrors (OAPs) typically used to direct and focus THz radiation.  We demonstrate, through simulations and data, that our proposed modified mirror geometry significantly reduces optical aberrations, which stem from finite sized source effects \cite{Bruckner2010, Arguijo:03}, resulting in a substantial reduction in the THz focal spot size and thus easing the requirement of large sample sizes in TDTS spectroscopy.

\begin{figure*}[tb]
\includegraphics[width=1.0\columnwidth, height=3.25in]{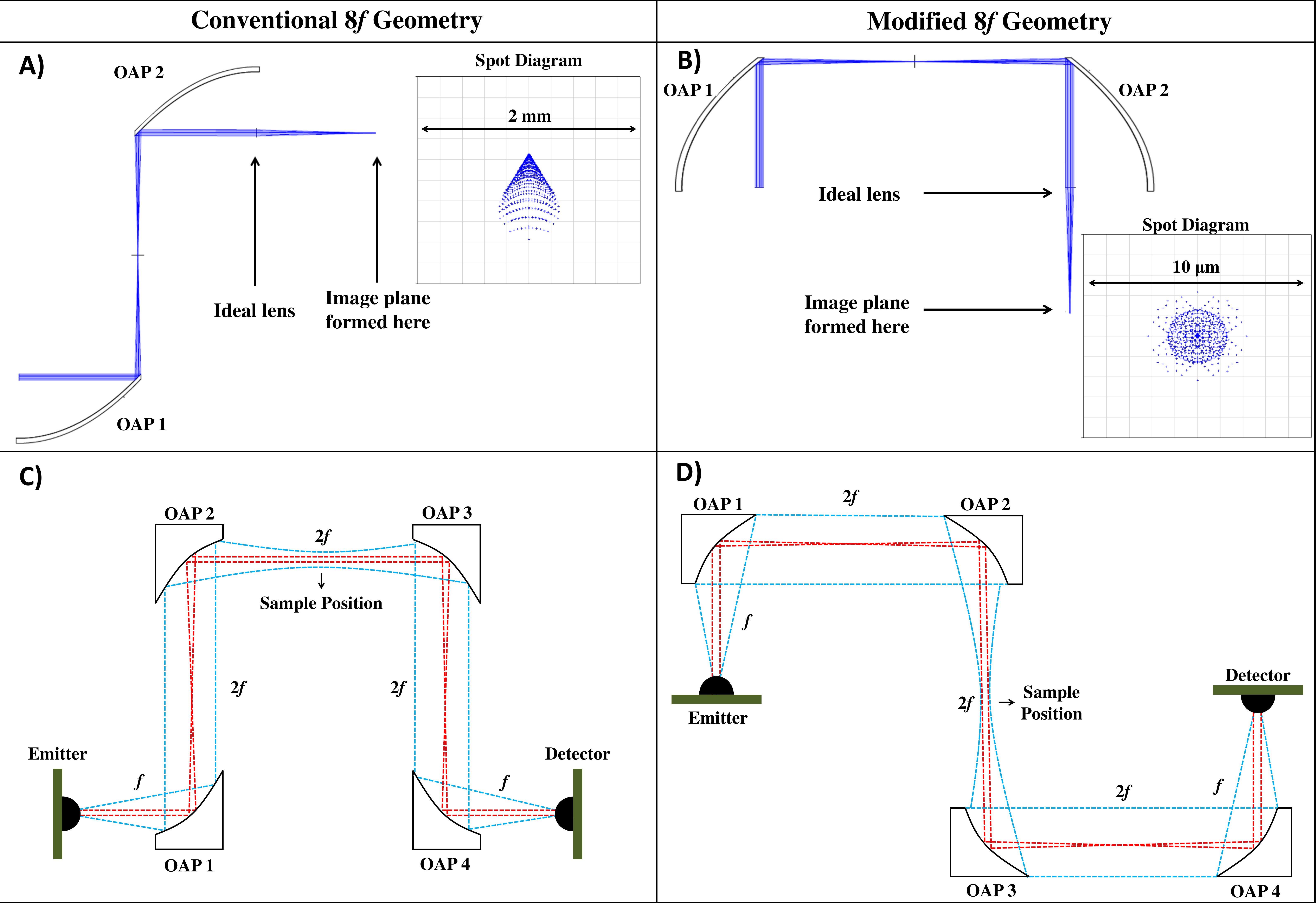}
\caption{Diagram of our ray tracing analysis, performed in the optical design software ZEMAX, for both the conventional (A) and modified (B) geometries.  Collimated beams were sent into OAP 1 to simulate the behavior of the high frequency light in the $8f$ geometry.  A significant coma is seen in the image formed in the conventional geometry while the image in the modified geometry is symmetric, diffraction limited, and greatly reduced in size.  Note that the scale of the spot diagram in the modified geometry is two orders of magnitude smaller than that of the conventional geometry. (C) Schematic of a TDTS spectrometer designed in the conventional $8f$ geometry as compared to (D) the modified $8f$ geometry.  The behavior of the low (blue lines) and high (red lines) frequencies in the $8f$ geometry are shown.}
\label{Fig1}
\end{figure*}

\section{Theory}

Off-axis parabolic mirrors are beneficial for imaging in the sub-millimeter range due to their achromatic and nearly lossless reflection properties.  However, previous theoretical classical aberration analyses and ZEMAX optical design software simulations have shown that the reduction of symmetry in their off-axis design introduces optical aberrations in their imaging capabilities \cite{Bruckner2010, Arguijo:03}.  It is known that OAPs suffer from two forms of optical aberration, coma and astigmatism, the ratio of which depends on geometrical properties of the OAP.  However, combinations of OAPs can compensate for these optical aberrations.  For instance, Br{\"u}ckner et al.\cite{Bruckner2010} found through numerical simulations that the imaging capability of a set of OAPs depends greatly on the alignment of one OAP to the next. While no alignment of two OAPs resulted in a complete cancellation of all optical aberrations, it was found that certain alignments perform significantly better than others. For instance, the 180 degree alignment of two OAPs, where the second OAP is a 180 degree rotation of the first, resulted in an enhancement of the optical aberrations of the two OAPs and correspondingly the worst imaging capabilities of all alignments studied.  We refer to this alignment as the ``conventional geometry" as it is often the alignment upon which TDTS spectrometers are based.  However, it was found that if the second OAP is aligned such that it is the mirror image of the first OAP, in which case the two parent parabolas share an axis, then the principle aberrations of the first OAP are compensated by the second OAP, resulting in a diffraction limited image.  This alignment produced the best imaging results.  We correspondingly refer to this alignment as the ``modified geometry." Fig. \ref{Fig1} (A) \& (B) show two OAPs aligned in the conventional and modified geometries respectively with some further analysis which will be discussed below.

Experimental configurations for directing the THz radiation onto a sample under test are based on a related alignment scheme of OAPs in what is known as the ``$8f$ configuration." This geometry, in which the total THz beam path is 8 times the reflected focal length of one OAP, allows for the passage of both low and high frequencies through the system in a consistent fashion.  One can gain some understanding of wave propagation through this system from a ray optics perspective.  Fig. \ref{Fig1} (C) \& (D) show the beam path of both the low (blue lines) and high (red lines) frequencies in a ray optics perspective for spectrometers based on the conventional and modified geometries respectively.  The THz emitter is placed at the focal spot of a 90 degree off-axis parabolic mirror, OAP 1, of reflected focal length $f$. A collimating hyper-hemispherical lens is placed directly after the emitter to correct finite size source effects and to more efficiently couple THz radiation to free space.  In the low frequency limit, the wave fronts emerge from the hyper-hemispherical lens and quickly diverge in an approximately Gaussian manner before OAP 1.  OAP 1 then collimates the low frequencies onto OAP 2 which then focuses the light halfway between OAPs 2 and 3.  The high frequencies behave in the reverse manner as they emerge from the hyper-hemispherical lens collimated, are focused by OAP 1, and collimated again by OAP 2.  This results in a near frequency independent focus, or beam waist, formed between OAPs 2 and 3 where the sample under test is placed.  The second half of the system is simply the inverse of the first half as shown in Fig. \ref{Fig1} (C) \& (D). This analysis provides perspective on the light propagation through the system, but does not give insight into the above described optical aberrations arising from finite sized source effects.

\begin{figure}
\includegraphics[width=0.75\columnwidth, height=8in]{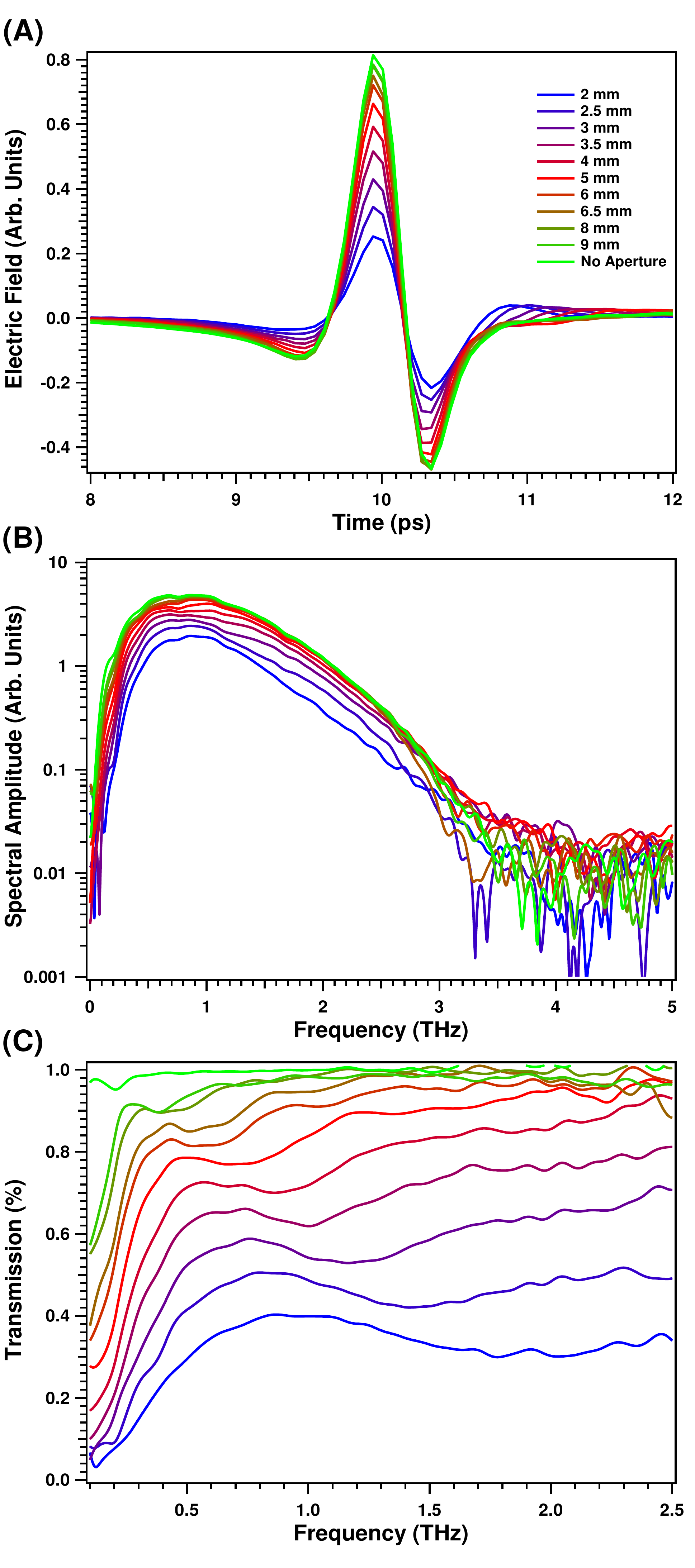}
\caption{(A) Electric field plotted as a function of time for different sized apertures placed at the focal spot. (B) Corresponding Fourier transforms of the electric fields shown in (A) plotted on log scale. (C) Transmissions for each aperture size, defined as the ratio of the Fourier transform of the transmitted electric field through each aperture size divided by the Fourier transform of the electric field of a scan with no aperture present.}
\label{Fig2}
\end{figure}

\section{Methods}

The analysis of Br{\"u}ckner et al.\cite{Bruckner2010} showed the modified geometry has reduced optical aberrations when light is sent into OAP 1 diverging from the focus. As discussed above, this is only consistent with the behavior of the low frequencies in the $8f$ geometry.  The effects of optical aberrations for the high frequency light, which come into OAP 1 collimated (i.e. propagating into OAP 1 "backwards"), are unknown.  To investigate, we performed our own ZEMAX ray tracing simulations in which collimated light was sent into OAP 1.  Although the OAP, and all other optical elements in the simulation, are ideal, aberrations are still induced due to the backwards path of the light into the OAP.  An ideal lens was used to focus the light so that the imaging performance of both the conventional and modified OAP configurations could be studied. Spot diagrams were generated by projecting the image onto a plane placed at the focus of the ideal lens.

Fig. \ref{Fig1} (A) \& (B) show our ray tracing results for the conventional and modified geometries respectively.  We find the image in the conventional geometry to be highly distorted from coma, as seen in the spot diagram in Fig. \ref{Fig1} (A).  Additionally, we find the image size in this geometry to be approximately twice the image size formed by a single OAP, confirming that the aberrations are enhanced in this configuration.  However, the spot diagram for the modified geometry, Fig. \ref{Fig1} (B), shows the image to be undistorted and free of aberrations. It should be noted that the scale of the spot diagram in the modified geometry is two orders of magnitude smaller than that of the conventional geometry. This suggests that the optical aberrations in the high frequency light are also compensated in the modified geometry. Thus the modified geometry has significantly reduced optical aberrations for the entire range of frequencies present in a typical THz pulse as shown by both our and Br{\"u}ckner et al.'s \cite{Bruckner2010} ray tracing analysis.

As discussed above, TDTS spectrometers in the $8f$ configuration are composed of two copies of the OAP configurations shown in Figs. \ref{Fig1} (A) \& (B).  Figs. 1 (C) \& (D) show two $8f$ spectrometers comprised of the conventional and modified configurations shown in Figs. \ref{Fig1} (A) \& (B).  We refer to these as the ``conventional $8f$ geometry" and ``modified $8f$ geometry" respectively.  Both systems have a focal spot formed at the sample position between OAPs 2 and 3.  However, as shown above, the quality of that focal spot is determined by the relative alignment of OAPs 1 and 2 as well as OAPs 3 and 4.  One can see that the conventional geometry consists of the 180 degree alignment of OAPs 1 and 2 that was previously shown to produce the poorest image quality.  However, our modified geometry is based on the mirror image alignment of OAPs 1 and 2 that results in the compensation of the principle aberration and an ideally diffraction limited focal spot.  

To test our theoretical results, a home-built TDTS spectrometer was constructed in the modified $8f$ geometry as shown in Fig. \ref{Fig1} (D).  The system utilizes aluminum coated OAPs with a diameter and reflected focal length of 2 in and 4 in respectively.

\section{Results and Discussion}

Fig. \ref{Fig2} shows data taken on our spectrometer.  To investigate the imaging capabilities of the system, data was taken with an aperture placed at the focal spot, the size of which was then varied.  The apertures used were metallic circular apertures attached to a mount with $\hat{x}$, $\hat{y}$, and $\hat{z}$ axis adjustment.  The focal spot was defined as the position in which the maximum amount of signal was transmitted through the smallest aperture.  Fig. \ref{Fig2} (A) displays the measured electric field of the THz pulse as a function of time for each aperture studied, including a scan in which no aperture was present. Fig. \ref{Fig2} (B) shows the corresponding magnitude of the Fourier transforms of the pulses found in Fig. \ref{Fig2} (A) on log scale. Fig. \ref{Fig2} (C) shows the magnitude of the complex transmission for each aperture size, defined as the ratio of the Fourier transform of the transmitted electric field through each aperture size divided by the Fourier transform of the electric field of a scan with no aperture present.

These data show an expected monotonic decrease in both the time domain and the frequency domain in the magnitude of the signal as the aperture size is reduced.  However, we find that even for the smallest aperture size, 2 mm, that a substantial amount of light is transmitted, with the maximum transmission being slightly greater than 40\% at $\approx$ 0.9 THz.  Furthermore, the measured signal of the 9 mm aperture is nearly indistinguishable from data for which no aperture is present, especially above $\approx$ 0.3 THz.  More can be learned from the frequency dependence of the transmission.  The transmissions decrease dramatically below a low frequency cut off.  This cut off, whose frequency decreases with increasing aperture size, can be explained by simple far field diffraction of light.  Above this cut off we see a maximum in the transmission followed by a region of slowly increasing transmission with frequency.  This is consistent with an approximately Gaussian beam incident on a circular aperture \cite{Uehara:86}.

For comparison, an identical aperture experiment was performed on a separate home-built THz spectrometer designed in the conventional $8f$ geometry.  While this conventional system utilizes OAPs with slightly longer reflected focal lengths, 6 inches, we believe it to be fairly representative of the conventional $8f$ geometry.

We define the integral over all frequency of the magnitude of the Fourier transforms squared as a figure of merit to characterize each system.  This quantity is identical to the integral of the electric field squared as a function of time, Fig. \ref{Fig2} (A), via Parseval's theorem, and is therefore proportional to the total power contained in each THz pulse.  

\begin{equation}
P_{total} \propto \int_{-\infty}^{\infty} |E(t)|^2 dt = \int_{-\infty}^{\infty} |E(\omega)|^2 d \omega 
\label{eq1}
\end{equation}

\noindent This figure of merit is chosen because it allows us to quantify the focal spot size of both TDTS systems.  The transmitted power $P(r, z)$ of a Gaussian beam incident on a circular aperture of diameter $d$ at position $z$ is given by the equation \cite{Tanaka:85}, 
\begin{equation}
P(r, z) = P_0[1 - e^{-2d^2/w(z)^2}].
\label{eq2}
\end{equation}

\noindent In this expression $P_0$ represents the total power, i.e. the power with no aperture is present, and $w(z)$ is beam waist diameter at position $z$.  With an aperture placed at the focal position, $w(z)$ is a constant and represents the focal spot diameter. Thus by fitting our figure of merit to Eq. \ref{eq2}, the focal spot diameter for our TDTS spectrometers in both the conventional and modified $8f$ geometries can be extracted.  

\begin{figure}[tb]
\includegraphics[width=0.75\columnwidth, height=2.5in]{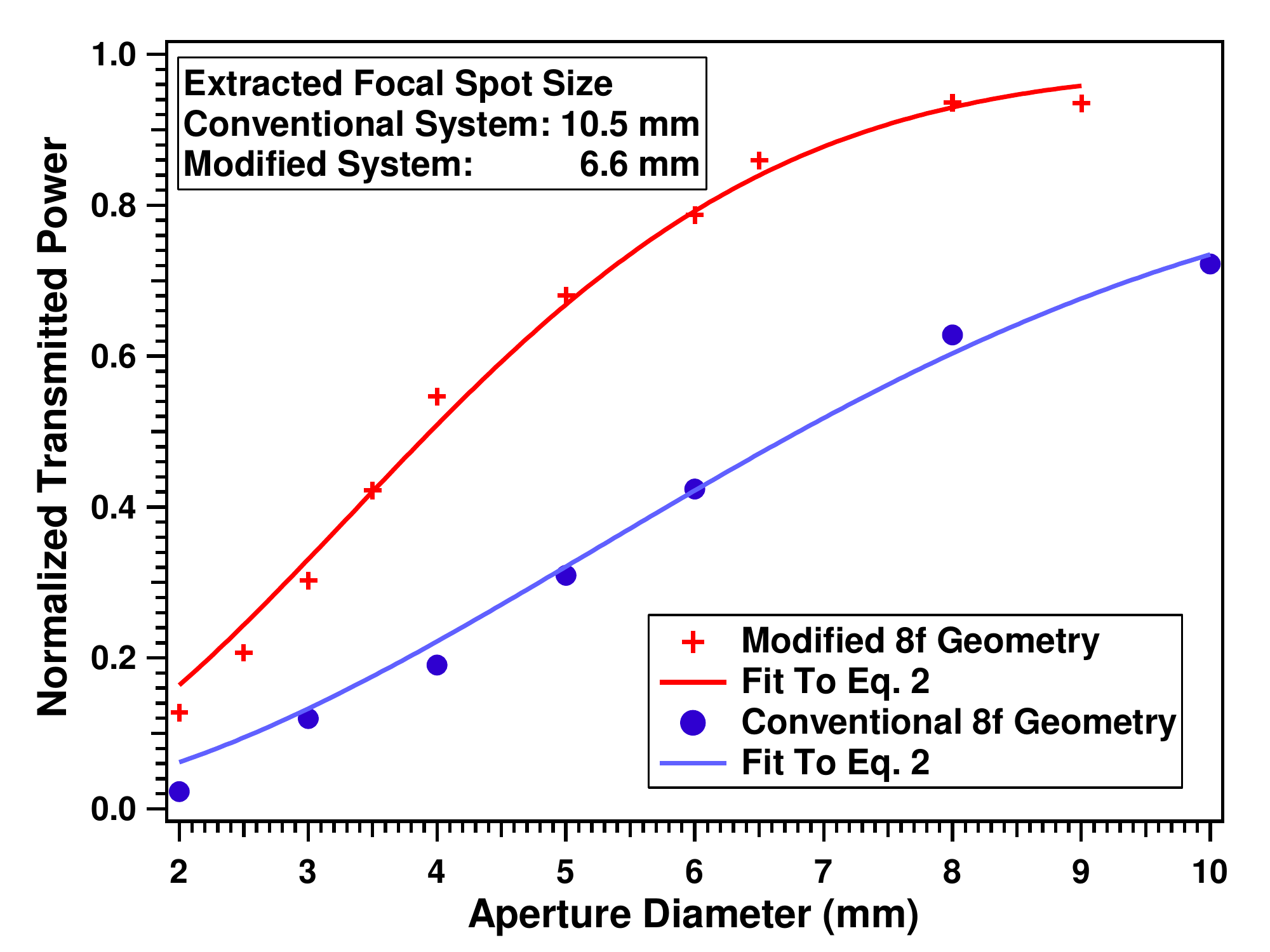}
\caption{Total transmitted power as a function of aperture size for TDTS systems designed in both the modified (red, crosses) and conventional (blue, circles) configurations.  Solid lines are fits of the data to Eq. \ref{eq2} as discussed in the text.  The extracted focal spot sizes for the conventional and modified geometries are 10.5 mm and 6.6 mm respectively.}
\label{Fig3}
\end{figure}

Fig. \ref{Fig3} shows the normalized transmitted power as a function of aperture size for both the modified $8f$ (red, crosses) and conventional $8f$ (blue, circles) geometries.  Normalization was performed by dividing the transmitted power of each aperture size by that of the transmitted power with no aperture present.  In this way, we remove any systematic differences between the two TDTS systems.  One can immediately see from Fig. 3 that the modified $8f$ geometry results in a dramatic improvement for all aperture sizes studied.  We find more than double the signal in the modified $8f$ geometry than in the conventional $8f$ geometry for aperture sizes less than 8 mm. More importantly, the best improvement over the conventional $8f$ system is found for the smallest apertures.  We find the transmitted power through a 2 mm aperture to be 4 times greater in the modified $8f$ geometry as compared to the conventional $8f$ geometry, greatly easing the typically constraining need for large sample sizes in TDTS.

Also shown in Fig. 3 are the fits for both the conventional $8f$ and modified $8f$ geometries to Eq. \ref{eq2}.  As explained above, the focal spot diameter can be extracted by fitting these data to Eq. \ref{eq2}.  We extract values of 10.5 mm and 6.6 mm for the conventional $8f$ and modified $8f$ geometries respectively. One should note that Eq. \ref{eq2} only applies for monochromatic Gaussian beams which, as shown by the ray tracing results of Fig. \ref{Fig1} (A), is not applicable to the conventional $8f$ system.  However, we believe the larger extracted focal spot diameter is a phenomenological description of the conventional $8f$ system and still emblematic of the larger optical aberrations in this geometry.

\section{Conclusion}

We have demonstrated that a simple rearrangement of the off-axis parabolic mirrors results in a significantly reduced focal spot size and correspondingly improved performance of time domain terahertz spectrometers.  We demonstrated that this improvement is caused by reduced optical aberrations in our ``modified $8f$ geometry" as compared to systems designed in the ``conventional $8f$ geometry."  We demonstrated through simulations that the aberrations in the modified geometry are compensated from one OAP to the next, instead of enhanced as they are in the conventional geometry.  We found that a system designed in the modified $8f$ geometry significantly out-performs a similar system designed in the conventional $8f$ geometry with the largest improvement seen at the smallest aperture size. We found a near 40\% reduction in THz focal spot size in the modified $8f$ geometry as compared to the conventional $8f$ system.  This design enables the study of smaller samples than what has previously been capable of in TDTS experiments.  \\

\begin{acknowledgements}
The THz instrumentation development was funded by the Gordon and Betty Moore Foundation through Grant GBMF2628 to NPA. The authors would like to thank C. M. Morris, M. Neshat, and LiDong Pan for helpful conversations.
\end{acknowledgements}

\bibliographystyle{unsrt}

\bibliography{BLCBib}

\end{document}